\documentclass[conference]{IEEEtran}
\IEEEoverridecommandlockouts

\usepackage{cite}
\usepackage{amsmath,amssymb,amsfonts}
\usepackage{algorithmic}
\usepackage{graphicx}
\usepackage{textcomp}
\usepackage{xcolor}
\usepackage{multirow}
\usepackage{float}
\usepackage{hyperref}
\usepackage{booktabs}
\usepackage{pifont}

\usepackage{colortbl} % Add this line in the preamble

\def\BibTeX{{\rm B\kern-.05em{\sc i\kern-.025em b}\kern-.08em
    T\kern-.1667em\lower.7ex\hbox{E}\kern-.125emX}}
\begin{document}

\title{Towards Cross-Modal Text-Molecule Retrieval with Better Modality Alignment\\
}

\author{\IEEEauthorblockN{Jia Song\textsuperscript{1}, Wanru Zhuang\textsuperscript{2}, Yujie Lin\textsuperscript{2}, Liang Zhang\textsuperscript{2}, Chunyan Li\textsuperscript{3}, Jinsong Su\textsuperscript{1,2$\dagger$}, Song He\textsuperscript{4$\dagger$}, Xiaochen Bo\textsuperscript{4$\dagger$}}
\IEEEauthorblockA{
\textsuperscript{1}\textit{Institute of Artificial Intelligence, Xiamen University, Xiamen, China} \\
\textsuperscript{2}\textit{School of Informatics, Xiamen University, Xiamen, China}\\
\textsuperscript{3}\textit{School of Informatics, Yunnan Normal University, Kunming, China}\\
\textsuperscript{4}\textit{Institute of Health Service and Transfusion Medicine, Beijing, China}\\
\{songjia, zwramy, linyj, lzhang\}@stu.xmu.edu.cn, 
lchy0316@gmail.com, \
jssu@xmu.edu.cn, \{hes1224, boxiaoc\}@163.com}
\thanks{$\dagger$ Corresponding authors.}
}

\maketitle

\begin{abstract}

Cross-modal text-molecule retrieval model aims to learn a shared feature space of the text and molecule modalities for accurate similarity calculation, which facilitates the rapid screening of molecules with specific properties and activities in drug design.
    However, previous works have two main defects.
    First, they are inadequate in capturing modality-shared features considering the significant gap between text sequences and molecule graphs.
    Second, they mainly rely on contrastive learning and adversarial training for cross-modality alignment, both of which mainly focus on the first-order similarity, ignoring the second-order similarity that can capture more structural information in the embedding space.
    To address these issues, we propose a novel cross-modal text-molecule retrieval model with two-fold improvements.
    Specifically, on the top of two modality-specific encoders, we stack a memory bank based feature projector that contain learnable memory vectors to extract modality-shared features better.
    More importantly, during the model training, we calculate four kinds of similarity distributions (text-to-text, text-to-molecule, molecule-to-molecule, and molecule-to-text similarity distributions) for each instance, and then minimize the distance between these similarity distributions (namely second-order similarity losses) to enhance cross-modal alignment.
    Experimental results and analysis strongly demonstrate the effectiveness of our model.
    Particularly, our model achieves SOTA performance, outperforming the previously-reported best result by 6.4\%.

\end{abstract}


\section{Introduction}

Traditional drug development is a long process, where pharmacologists often retrieve existing databases for better molecule design or investigation of newly-discovered compounds.
Prominent databases such as PubChem~\cite{kim2016pubchem,kim2019pubchem} and DrugBank~\cite{kratochvil2019interoperabledrugbank} contain large-scale structures and properties of known molecules, which foster the design of novel molecules with specific biological activities.
However, most existing retrieval systems are based on uni-modal methods~\cite{swamidass2007mathematical,baldi2007lossless,tang2012functional,gadaleta2018new,qu2018pharmki,intarapaiboon2011extracting}, where we must first conduct a uni-modal search and then identify the cross-modal counterpart of the retrieved result.
Unfortunately, a considerable proportion of these molecules are not covered by expert-annotated text descriptions, which result in unsatisfactory retrieval results attributed to the scarcity of paired data.
Hence, it is of significance to develop a cross-modal text-molecule retrieval model, which enables pharmacologists to access compound information more efficiently.

In recent years, with the rapid development of deep learning, neural network based cross-modal retrieval models for drug design have attracted much attention.
In this aspect, KV-PLM \cite{kvplm} inserts SMILES sequences of molecules into paired text for pre-training.
Both text and SMILES sequences are encoded using the same encoder, facilitating implicit modality alignment.
Meanwhile, more researchers use molecule graph based GNNs \cite{momusu,liu2022multimoleculestm,zhao2023adversarial}, such as Text2Mol \cite{edwards2021text2mol}, to obtain more informative molecule representations.
Generally, these studies first employ two pre-trained encoders to separately encode the input text and molecule graph, and then apply contrastive learning to achieve cross-modality alignment.
Along this line, \cite{zhao2023adversarial} extend Text2Mol to AMAN,
which additionally introduces adversarial training to conduct modality alignment, setting a new state-of-the-art (SOTA).

However, previous works still exhibit limitations in modality alignment due to the following two reasons: 
1) there is no modal interaction observed after the encoders as reported in \cite{edwards2021text2mol,liu2022multimoleculestm,momusu}, or merely limited to a linear layer as in \cite{zhao2023adversarial}, which is inadequate in capturing modality-shared features due to significant differences between the topological structures and pre-trained feature spaces of text sequences and molecule graphs;
2) they only use the objectives based on first-order similarity, i.e., contrastive learning and adversarial training, ignoring the second-order similarity which captures rich structural information in the embedding space.

\begin{figure*}[ht]
\centering
\includegraphics[width=0.8\linewidth]{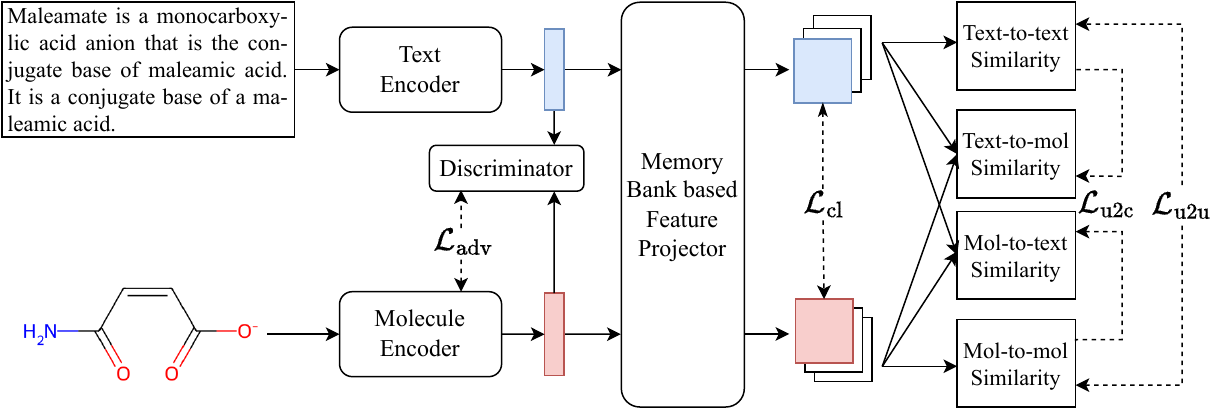}
% \vspace{-10pt}
\caption{The architecture of our model. It mainly consists of four modules: a text encoder, a molecule encoder and a discriminator distinguishing between two modalities, and a memory bank based feature projector that introduces learnable memory vectors to learn the multi-modal feature space. In addition to the conventional contrastive learning loss $\mathcal{L}_{cl}$ and adversarial training loss $\mathcal{L}_{adv}$, we incorporate second-order similarity losses $\mathcal{L}_{\mathrm{u2u}}$ and $\mathcal{L}_{\mathrm{u2c}}$ to enhance cross-modality alignment.}
\label{fig:paradigm}
\end{figure*}

In this paper, we propose a novel model for cross-modal text-molecule retrieval, which achieves better cross-modal alignment via introducing a memory bank based feature projector and second-similarity losses to learn modality-shared features.
We improve the previous works from the perspectives of model architecture and training.
Architecture wise, the memory bank based feature projector shared by two modalities serves as a bridge to help modality alignment.
The memory vectors in it are queries in a cross-attention module to extract fix-sized shared semantic features from two modalities, thereby mitigating the modality gap.

When training our model, in addition to first-order similarity losses (contrastive learning and adversarial training losses), we introduce two second-order similarity losses, $\mathcal{L}_{\mathrm{u2u}}$ and $\mathcal{L}_{\mathrm{u2c}}$, to further enhance cross-modality alignment.
The basic intuition behind it from the fact that the similarity between instances should remain consistent regardless of modalities.
To this end, based on the representations of different modalities, we calculate two uni-modal and two cross-modal similarity distributions for each instance in training batch.
Then, we minimize the KL divergence $\mathcal{L}_{\mathrm{u2u}}$ between two uni-modal similarity distributions, and the KL divergence $\mathcal{L}_{\mathrm{u2c}}$ of uni-modal similarity distributions from cross-modal similarity distributions.
Note that, we do not use cross-modal similarity distributions to supervise the training of uni-modal similarity distributions.
This is because the uni-modal representations have been pre-trained on a large-scale data, their similarity distributions are more reliable.

Our second-order similarity losses consider neighborhood structures in the modality-shared semantic space, and thus are supplement to the conventional first-order similarity losses, which are contrastive learning and adversarial training losses based only on the representations of instances in different modalities.
Specifically, $\mathcal{L}_{\mathrm{u2u}}$ captures the neighboring information consistency in different modality feature spaces to enhance the alignment.
Unlike conventional contrastive learning using binary labels indicating whether the considered text and molecule are from the same instance, $\mathcal{L}_{\mathrm{u2c}}$ uses the uni-modal similarity distribution as a smoothing supervisory signal to learn the cross-modal similarity distribution.

Our contributions are summarized as follows:

\begin{itemize}
\setlength{\itemsep}{2pt}
\setlength{\parsep}{2pt}
\setlength{\parskip}{2pt}
\item We propose a memory bank-based feature projector that contains
learnable memory vectors to extract modality-shared features better, bridging the modality gap between text sequences and molecule graphs.
\item We propose several second-order similarity losses to enhance cross-modality alignment further. To the best of knowledge, previous studies only consider the first-order feature, but ignore the second-order neighbor relationships between instances.
\item Extensive experiments on ChEBI-20 and PCdes datasets strongly demonstrate the effectiveness and generalizability of our proposed model.
\end{itemize}

\section{Our Model}

As illustrated in Figure~\ref{fig:paradigm}, our model consists of two individual encoders responsible for encoding the input text and molecule, respectively.
On the top of encoders, a memory bank based feature projector is stacked to extract modality-shared features.
Besides, a discriminator is equipped for adversarial training, which is beneficial for cross-modality alignment, as analyzed in \cite{zhao2023adversarial}.
In the following, we first elaborate on different components of our model, and then give a detailed description of the model training.

\subsection{Model Architecture}

\paragraph{Encoders}

As a common practice, we use SciBERT \cite{scibertBeltagyLC19} stacked with a linear layer as our text encoder.
With this encoder, we derive a sequence of token representations $\textbf{H}^t$=$\{\textbf{h}_i^t\}^L_{i=1}$ from the given text, where $L$ is the total number of tokens.

Similar to Text2Mol \cite{edwards2021text2mol}, our molecule encoder is a graph convolutional network (GCN) \cite{iclrKipfW17gcn} stacked with a linear layer.
For each atom of the given molecule, we first initialize its representation with Mol2vec \cite{jaeger2018mol2vec}.
Then, we use the molecule encoder to learn the atom representations $\textbf{H}^m$=$\{\textbf{h}_i^m\}^K_{i=1}$, where $K$ is the total number of atoms in the molecule.

\paragraph{Memory Bank Based Feature Projector}

Due to the substantial disparities in the topological structures and pre-trained feature spaces of text and molecule, we introduce a memory bank based feature projector mapping both $\textbf{H}^t$ and $\textbf{H}^m$ to fixed-size vectors in a same semantic space, effectively bridging the gap between different modalities.

\begin{figure}[t]
\centering
\includegraphics[width=1\linewidth]{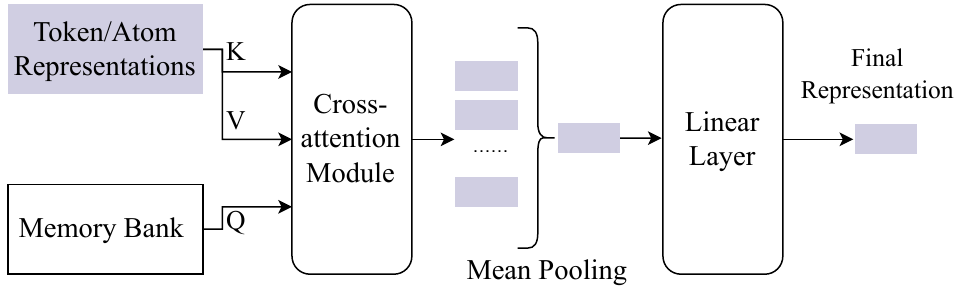}
\caption{The diagram illustrates how the outputs of the two encoders are processed by the memory bank based feature projector to obtain the final modality-shared feature representations.}
\label{fig:query}
\end{figure}

As shown in Figure~\ref{fig:query}, this projector consists of a memory bank, a cross-attention module and a linear layer.
In the memory bank, 
we introduce $n$ learnable memory vectors to extract modality-shared semantic information, where these vectors serve as the queries of cross-attention to interact with the token or atom representations.

Here, we take the text modality as example.
The projector takes the token representations $\textbf{H}^t$ output from the text encoder as input and then projects them to $n$ modality-shared feature vectors:
\begin{gather}
    \textbf{O}^t=\mathrm{Attn}(\textbf{Q}, \textbf{H}^t\textbf{W}_K, \textbf{H}^t\textbf{W}_V), 
\end{gather}
\noindent where $\textbf{Q}\in\mathbb{R}^{n\times d}$ is a matrix consisting of the above $n$ learnable memory vectors,
$\textbf{W}_K$ and $\textbf{W}_V$ are parameter matrices projecting $\textbf{H}^t$ to attention keys and values, respectively, and $\textbf{O}^t\in\mathbb{R}^{n\times d}$ denotes the output modality-shared feature matrix.

After applying mean pooling followed by a linear layer, we can obtain the final representation $\textbf{x}^t$ of text modality:
\begin{gather}
    \textbf{x}^t=\mathrm{FC}(\mathrm{meanPool}(\textbf{O}^t)).
\end{gather}
Similarly, we can obtain the final representation $\textbf{x}^m$ of the molecule by feeding $\textbf{H}^m$ into this projector.
Finally, we perform cross-modal retrieval with $\textbf{x}^t$ and $\textbf{x}^m$.

\subsection{Training Objective}

The training objective of our model involves three kinds of losses, including two first-order similarity losses: contrastive loss $\mathcal{L}_{\mathrm{cl}}$ and adversarial training loss $\mathcal{L}_{\mathrm{adv}}$, and two second-order similarity losses: $\mathcal{L}_{\mathrm{u2u}}$ and $\mathcal{L}_{\mathrm{u2c}}$.
Formally, the overall training objective is defined as
\begin{equation}
\label{loss}
    \mathcal{L}=\mathcal{L}_{\mathrm{cl}}+\lambda_1\mathcal{L}_{\mathrm{adv}}+\mathcal{L}_{\mathrm{u2u}}+\mathcal{L}_{\mathrm{u2c}},
\end{equation}
\noindent where $\lambda_1$ is a hyperparameter used to balance the training losses.
Notice that introducing more hyperparameters may potentially enhance the efficacy of other losses.
However, given the potential difficulties in model tuning with too many hyperparameters, we simply set the weights of other losses to 1, which has already yielded significant results. In Section~\ref{hyper}, we investigate the impact of the hyperparameters of $\mathcal{L}_{\mathrm{u2u}}$ and $\mathcal{L}_{\mathrm{u2c}}$ on the model performance.

\paragraph{First-order Similarity  Losses $\mathcal{L}_{\mathrm{cl}}$ and $\mathcal{L}_{\mathrm{adv}}$} 

The first-order similarity losses only consider the modality-specific representations of instances to achieve cross-modal alignment.
Specifically, contrastive learning aims to reduce the distance between the anchor-positive representations while increasing the distance between the anchor-negative ones.
Besides, adversarial training enables the encoders to generate similar representations for different modalities of instances.

As implemented in AMAN \cite{zhao2023adversarial}, we employ triplet contrastive learning to train our model.
To this end, we first construct training triplets, each of which comprises an anchor in one modality, a positive instance, and a negative instance from the other modality.
Then, we design $\mathcal{L}_{\mathrm{cl}}$ to increase the similarity between the anchor and the positive instance, while decreasing the similarity between the anchor and the negative instance simultaneously:

\begin{equation}
\begin{aligned}
\mathcal{L}_{\mathrm{cl}}= 
& \max (d(\textbf{x}^t_a, \textbf{x}_{p}^m)-d(\textbf{x}^t_a, \textbf{x}^m_{n})+\alpha, 0) \\\
&+\max (d(\textbf{x}^m_a, \textbf{x}^t_{p})-d(\textbf{x}^m_a, \textbf{x}^t_{n})+\alpha, 0),
\end{aligned}
\end{equation}

\noindent where $d(\cdot)$ is the Euclidean distance, $\alpha$ is a margin, and $\textbf{x}^*_a$, $\textbf{x}^*_p$, $\textbf{x}^*_n$ refer to the representations of anchor, positive and negative representations, respectively.

As described above, our model is equipped with a discriminator $D$, which conducts adversarial training with the molecule encoder.
During the specific training process, we adopt the WGAN-GP method \cite{gulrajani2017improved} to perform adversarial training, where $\mathcal{L}_{\mathrm{adv}}$ is defined as
\begin{equation}
\begin{aligned}
\mathcal{L}_{\mathrm{adv}}= & \mathbb{E}_{\textbf{g}^t \sim p_{\text {t}}}\left[\log D(\textbf{g}^t)\right]+\mathbb{E}_{\textbf{g}^m \sim p_{\text {m}}}\left[\log \left(1-D(\textbf{g}^m)\right)\right].
\end{aligned}
\end{equation}
\noindent Here, $p_{t}$ and $p_{m}$ are text and molecule distributions.
$\textbf{g}^t$ and $\textbf{g}^m$ are the global text and molecule representations, which are obtained by performing mean pooling on $\textbf{H}^t$ and $\textbf{H}^m$ from encoders, respectively.
During model training, we solve for the parameters of the molecule encoder and discriminator by a min-max optimization approach.

\paragraph{Second-order Similarity Losses $\mathcal{L}_{\mathrm{u2u}}$ and $\mathcal{L}_{\mathrm{u2c}}$}

In previous studies \cite{edwards2021text2mol,zhao2023adversarial}, contrastive learning and adversarial training are based on the learned representations of a single instance, while ignoring the similarities between different instances.
To tackle this issue, we propose two second-order similarity losses to benefit cross-modality alignment.

Concretely, given a training batch $B$ consisting of text-molecule pairs, 
we first compute four kinds of similarities between instances based on their text and molecule representations.
For example,
for the $i$-th and $j$-th instances in $B$, we calculate cosine distance between $\textbf{x}_i^t$ and $\textbf{x}_j^t$ as their text-to-text similarity, which are the text representations of the $i$-th and $j$-th instances, respectively.
Likewise, we compute their molecule-to-molecule similarity with $\textbf{x}_i^m$ and $\textbf{x}_j^m$, text-to-molecule similarity with $\textbf{x}_i^t$ and $\textbf{x}_j^m$, and molecule-to-text similarity based on $\textbf{x}_i^m$ and $\textbf{x}_j^t$.

Furthermore, 
we obtain four kinds of similarity distributions for each instance $i$:
text-to-text similarity distribution (TTSD) $\mathcal{P}_{i,:}^{tt}$,
molecule-to-molecule similarity distribution (MMSD) $\mathcal{P}_{i,:}^{mm}$,
text-to-molecule similarity distribution (TMSD) $\mathcal{P}_{i,:}^{tm}$,
and
molecule-to-text similarity distribution (MTSD) $\mathcal{P}_{i,:}^{mt}$.
Take $\mathcal{P}_{i,:}^{tt}=\{\mathcal{P}_{ij}^{tt}\}, \forall j \in B$ as example, it is calculated as
% \begin{equation}
% \label{p}
% \mathcal{P}_{i j}^{tt}=\frac{\exp \left(d(\textbf{x}_i^t, \textbf{x}_j^t)/\tau\right)}{\sum_{j'=1}^{|B|} \exp \left(d(\textbf{x}_i^t, \textbf{x}_{j'}^t)/\tau\right)}, \forall j \in B.
% \end{equation}
\begin{equation}
\label{p}
\mathcal{P}_{i j}^{tt}=\frac{\exp \left(d(\textbf{x}_i^t, \textbf{x}_j^t)\right)}{\sum_{j'=1}^{|B|} \exp \left(d(\textbf{x}_i^t, \textbf{x}_{j'}^t)\right)}, \forall j \in B.
\end{equation}

\begin{table*}[!ht]
\renewcommand{\arraystretch}{1.1}
    \centering
    \caption{Results of various models on the test sets on the ChEBI-20 dataset. The best results are marked in bold. $^\dag$: Please note that these results are directly cited from the previous studies~\cite{edwards2021text2mol,zhao2023adversarial}.}
    % \resizebox{0.9\linewidth}{!}{
    \begin{tabular}{c|cccc|cccc}
    \toprule
        \multirow{2}{*}{\textbf{Models}} & \multicolumn{4}{c|}{\textbf{Text-Molecule Retrieval}} & \multicolumn{4}{c}{\textbf{Molecule-Text Retrieval}} \\
        \cmidrule(lr){2-5} \cmidrule(lr){6-9} & \textbf{Hits@1($\uparrow$)} & \textbf{Hits@10($\uparrow$)}
        & \textbf{MRR($\uparrow$)} & \textbf{Mean Rank($\downarrow$)} & \textbf{Hits@1($\uparrow$)} & \textbf{Hits@10($\uparrow$)} & \textbf{MRR($\uparrow$)} & \textbf{Mean Rank($\downarrow$)} \\ 
        \midrule
        MLP-Ensemble$^\dag$~\cite{edwards2021text2mol} & 29.4\% & 77.6\% & 0.452 & 20.78 & - & - & - & - \\
        GCN-Ensemble$^\dag$~\cite{edwards2021text2mol} & 29.4\% & 77.1\% & 0.447 & 28.77 & - & - & - & - \\
        All-Ensemble$^\dag$~\cite{edwards2021text2mol} & 34.4\% & 81.1\% & 0.499 & 20.21 & 25.2\% & 74.1\% & 0.408 & 21.77 \\
        \midrule
        MLP+Atten$^\dag$~\cite{edwards2021text2mol} & 22.8\% & 68.7\% & 0.375 & 30.37 & - & - & - & - \\
        MLP+FPG$^\dag$~\cite{edwards2021text2mol} & 22.6\% & 68.6\% & 0.374 & 30.37 & - & - & - & - \\
        \midrule
        Atomas-base~\cite{atomas2024zhang} & 50.1\% & 92.1\% & 0.653 & 14.49 & 45.6\% & 90.3\% & 0.614 & 15.12 \\
        AMAN$^\dag$~\cite{zhao2023adversarial} & 49.4\% & 92.1\% & 0.647 & 16.01 & 46.6\% & 91.6\% & 0.625 & 16.50 \\
        AMAN(GTN$\rightarrow$GCN) & 49.0\% & 90.2\% & 0.640 & 18.21 & 45.1\% & 90.1\% & 0.605 & 17.55 \\
        
        \midrule
        \textbf{Ours} & \textbf{56.5\%} & \textbf{94.1\%} & \textbf{0.702} & \textbf{12.66} & \textbf{52.3\%} & \textbf{93.3\%} & \textbf{0.673} & \textbf{12.29} \\
    \bottomrule
    \end{tabular}
    % }
    \label{tab:main}
\end{table*}

\begin{table*}[ht!]
\renewcommand{\arraystretch}{1.1}
  \caption{Results of ablation study on the ChEBI-20 test set. $\mathcal{L}_{\mathrm{u2u}}$ consists of $\mathcal{L}_{\mathrm{t2m}}$ and $\mathcal{L}_{\mathrm{m2t}}$, which indicate the KL divergence of TTSD from MMSD and of MMSD from TTSD. $\mathcal{L}_{\mathrm{u2c}}$ consists of $\mathcal{L}_{\mathrm{2m}}$ and $\mathcal{L}_{\mathrm{2t}}$, which represent KL divergence of TMSD from MMSD and of MTSD from TTSD. MB is our learnable memory bank.}
  % \vspace{-15pt}
  \centering
    \resizebox{1\textwidth}{!}{
    \begin{tabular}{rccccrrrrrrrrrr}
          &       &       &       &       &       &       &       &       &       &       &       &       &       &  \\
\cmidrule{2-15}          & \multirow{2}[2]{*}{\#} & \multirow{2}[2]{*}{$\mathcal{L}_{t2m}$} & \multirow{2}[2]{*}{$\mathcal{L}_{m2t}$} & \multirow{2}[2]{*}{$\mathcal{L}_{2m}$} & \multicolumn{1}{c}{\multirow{2}[2]{*}{$\mathcal{L}_{2t}$}} & \multicolumn{1}{c|}{\multirow{2}[2]{*}{MB}} & \multicolumn{4}{c|}{Text-to-Molecule Retrieval} & \multicolumn{4}{c}{Molecule-to-Text  Retrieval} \\
          &       &       &       &       &       & \multicolumn{1}{c|}{} & \multicolumn{1}{c}{Mean Rank} & \multicolumn{1}{c}{MRR} & \multicolumn{1}{c}{Hits@1} & \multicolumn{1}{c|}{Hits@10} & \multicolumn{1}{c}{Mean Rank} & \multicolumn{1}{c}{MRR} & \multicolumn{1}{c}{Hits@1} & \multicolumn{1}{c}{Hits@10} \\
\cmidrule{2-15}          & 1     &\checkmark&    \checkmark   &      \checkmark &  \checkmark     & \multicolumn{1}{c|}{\checkmark} & \multicolumn{1}{c}{\textbf{12.66}} & \multicolumn{1}{c}{\textbf{0.702}} & \multicolumn{1}{c}{\textbf{56.5\%}} & \multicolumn{1}{c|}{\textbf{94.1\%}} & \multicolumn{1}{c}{\textbf{12.29}} & \multicolumn{1}{c}{\textbf{0.673}} & \multicolumn{1}{c}{\textbf{52.3\%}} & \multicolumn{1}{c}{\textbf{93.3\%}} \\
          & 2     &&\checkmark   &\checkmark&\checkmark&\multicolumn{1}{c|}{\checkmark} & \multicolumn{1}{c}{14.86} & \multicolumn{1}{c}{0.688} & \multicolumn{1}{c}{54.6\%} & \multicolumn{1}{c|}{93.7\%} & \multicolumn{1}{c}{14.62} & \multicolumn{1}{c}{0.665} & \multicolumn{1}{c}{51.4\%} & \multicolumn{1}{c}{92.9\%} \\
          
          & 3     &    &    &\checkmark&\checkmark& \multicolumn{1}{c|}{\checkmark} & \multicolumn{1}{c}{15.36} & \multicolumn{1}{c}{0.681} & \multicolumn{1}{c}{53.8\%} & \multicolumn{1}{c|}{93.0\%} & \multicolumn{1}{c}{13.62} & \multicolumn{1}{c}{0.656} & \multicolumn{1}{c}{50.5\%} & \multicolumn{1}{c}{92.8\%} \\
          
          & 4     &    &    &  &\checkmark& \multicolumn{1}{c|}{\checkmark} & \multicolumn{1}{c}{17.62} & \multicolumn{1}{c}{0.671} & \multicolumn{1}{c}{52.8\%} & \multicolumn{1}{c|}{93.0\%} & \multicolumn{1}{c}{16.19} & \multicolumn{1}{c}{0.643} & \multicolumn{1}{c}{49.0\%} & \multicolumn{1}{c}{91.9\%} \\

          & 5     &   &   &   & \multicolumn{1}{c}{} & \multicolumn{1}{c|}{\checkmark} & \multicolumn{1}{c}{19.29} & \multicolumn{1}{c}{0.655} & \multicolumn{1}{c}{50.6\%} & \multicolumn{1}{c|}{92.4\%} & \multicolumn{1}{c}{17.60} & \multicolumn{1}{c}{0.632} & \multicolumn{1}{c}{47.6\%} & \multicolumn{1}{c}{91.8\%} \\
          
          & 6     &  &   &   & \multicolumn{1}{c}{} & \multicolumn{1}{c|}{} & \multicolumn{1}{c}{21.77} & \multicolumn{1}{c}{0.637} & \multicolumn{1}{c}{48.7\%} & \multicolumn{1}{c|}{90.2\%} & \multicolumn{1}{c}{20.01} & \multicolumn{1}{c}{0.617} & \multicolumn{1}{c}{45.6\%} & \multicolumn{1}{c}{90.4\%} \\
\cmidrule{2-15}          &       &       &       &       &       &       &       &       &       &       &       &       &       &  \\
    \end{tabular}
    }
  \label{tab:ablation}%
\end{table*}

Then,
as shown in Figure~\ref{fig:paradigm},
we design two kinds of KL losses to enhance the consistency of different similarity distributions, so as to benefit cross-modality alignment:
1)~$\mathcal{L}_{\mathrm{u2u}}$. Via this loss, we aim to bring the two \textbf{U}ni-modal similarity distributions (TTSD and MMSD) together bidirectionally;
2)~$\mathcal{L}_{\mathrm{u2c}}$. By minimizing this loss, we expect to transfer the knowledge from the \textbf{U}ni-modal similarity distributions (TTSD and MMSD) to the \textbf{C}ross-modal similarity distributions (MTSD and TMSD).
Formally,
we define $\mathcal{L}_{\mathrm{u2u}}$ as the KL divergence between TTSD and MMSD to directly alignment these second-order similarity distributions:
\begin{equation}
\begin{aligned}
\mathcal{L}_{\mathrm{u2u}}=\frac{1}{|B|} \sum_{i=1}^{|B|} \left(K L\left(\mathcal{P}_{i,:}^{tt} \| \mathcal{P}_{i,:}^{mm}\right) + KL\left(\mathcal{P}_{i,:}^{mm} \| \mathcal{P}_{i,:}^{tt}\right) \right).
\end{aligned}
\end{equation}

\noindent Meanwhile, $\mathcal{L}_{\mathrm{u2c}}$ is the KL divergence of MMSD from TMSD and TTSD from MTSD, which is formulated as
\begin{equation}
\begin{aligned}
\mathcal{L}_{\mathrm{u2c}}=\frac{1}{|B|} \sum_{i=1}^{|B|} \left(K L\left(\mathcal{P}_{i,:}^{tt} \| \mathcal{P}_{i,:}^{mt}\right) + KL\left(\mathcal{P}_{i,:}^{mm} \| \mathcal{P}_{i,:}^{tm}\right)\right).
\end{aligned}
\end{equation}

Note that the two encoders have been pre-trained in large-scale uni-modal data, making uni-modal similarity distribution more reliable than cross-modal similarity distribution. Therefore, we use the former as soft labels to guide the later.

\section{Experiments}
\subsection{Setup}

\paragraph{Dataset}
We conduct experiments on the ChEBI-20 \cite{edwards2021text2mol} dataset. 
It consists  of 33,010 compound-description pairs from PubChem \cite{kim2016pubchem} and Chemical Entities of Biological Interest (ChEBI) \cite{hastings2016chebi}.

\paragraph{Evaluation Metrics}
We evaluate our model on two tasks: text-to-molecule retrieval and molecule-to-text retrieval.
Consistent with prior studies \cite{edwards2021text2mol,zhao2023adversarial}, we evaluate results by searching all instances in the dataset.
Since the text and molecule in this dataset are one-to-one correspongding, we utilize Mean
Reciprocal Rank (MRR), Mean Rank (MR), Hits@1,
and Hits@10 as evaluation metrics.

\paragraph{Baselines}
To verify the superiority of our model, we compare it with the following state-of-the-art baselines.

\begin{itemize}
\setlength{\itemsep}{2pt}
\setlength{\parsep}{2pt}
\setlength{\parskip}{2pt}
\item \textbf{MLP-Ensemble}, \textbf{GCN-Ensemble}, \textbf{All-Ensemble}, \textbf{MLP+Atten} and \textbf{MLP+FPG}~\cite{edwards2021text2mol}. The ensemble baselines involving different models with MLP- or GCN-based molecule encoders, 
where each model is initialized with different parameters.
The MLP+Atten and MLP+FPG baselines adopt attention-based association rules and FPGrowth algorithm to rerank the retrieval results, respectively.
\item \textbf{Atomas}~\cite{atomas2024zhang} Atomas is pretrained on a large-scale dataset where the molecules are represented by SMILES, aligning the two modalities at three granularities.
We train Atomas on the ChEBI-20 dataset using the same configuration as ours to perform a fair comparsion.
\item \textbf{AMAN} \cite{zhao2023adversarial}. It employs GTN \cite{gtnShiHFZWS21} as the molecule encoder and SciBERT \cite{scibertBeltagyLC19} as the text encoder, with triplet contrastive learning and adversarial training losses as the training objectives. 
Since AMAN is not open source, we also report the performance of \textbf{AMAN(GTN$\rightarrow$ GCN)} for analysis, which utilizes GCN instead of GTN as the molecule encoder.
Note that both AMAN and AMAN(GTN$\rightarrow$GCN) are our most important baselines.
\end{itemize}

\begin{table}[t]
  \centering
  \caption{The results of bidirectional cross-modal retrieval tasks on the PCdes test set.}
    \resizebox{0.4\textwidth}{!}{\begin{tabular}{c|cc|cc}
    \toprule
    \multirow{3}[4]{*}{Model} & \multicolumn{2}{c|}{sentence-level} & \multicolumn{2}{c}{paragraph-level} \\
          & Acc   & Rec@20 & Acc   & Rec@20 \\
\cmidrule{2-5}          & \multicolumn{4}{c}{Molecule-to-Text Retrieval} \\
    \midrule
    Sci-BERT & 50.38 & 62.11 & 62.57 & 60.67 \\
    KV-PLM* & 55.92 & 68.59 & 77.92 & 75.93 \\
    % MoMu-K_paper & 58.74 & 81.29 & 81.09 & 80.15 \\
    MoMu-K & 57.80 & 81.52 & 83.32 & 82.82 \\
    Ours  & \textbf{64.09} & \textbf{82.55} & \textbf{88.60} & \textbf{87.62} \\
    \midrule
    Model & \multicolumn{4}{c}{Text-to-Molecule Retrieval} \\
    \midrule
    Sci-BERT & 50.12 & 68.02 & 61.75 & 60.77 \\
    KV-PLM* & 55.61 & 74.77 & 77.03 & 75.47 \\
    % MoMu-K_paper & 54.94 & \textbf{78.29} & 81.45 & 80.62 \\
    MoMu-K & 54.80 & \textbf{79.78} & 83.41 & 82.69 \\
    Ours  & \textbf{63.38} & 73.91 & \textbf{87.73} & \textbf{88.27} \\
    \bottomrule
    \end{tabular}}
      
  \label{tab:pcdes}%
\end{table}%

\paragraph{Implement Details}

We employ Mol2vec \cite{jaeger2018mol2vec} to preprocess the molecules in the training set with parameters the same as Text2Mol \cite{edwards2021text2mol}.
The text encoder is initialized with SciBERT, with the dimension of text representations projected from 768 to 300, aligned with the dimension of atom representations.
The memory vectors in our memory bank are 28 300-dimensional vectors.
The weight of $\mathcal{L}_{adv}$ is 2e-4, which is determined according to the model performance on validation set.
As implemented in AMAN \cite{zhao2023adversarial}, we set the margin of $\mathcal{L}_{cl}$ to 0.3.
We train the model for 60 epochs with a batch size of 32.
The model parameters are randomly initialized except of the two pre-trained encoders, and are updated using the Adam optimizer, with the learning rate set to 3e-5 for the text encoder and 1e-4 for other components of the model.

\subsection{Main Results}

Table \ref{tab:main} shows the experiment results on the test set of CheEBI-20 dataset for both text-to-molecule retrieval and molecule-to-text retrieval tasks.
Our model consistently outperforms all baselines across all evaluation metrics.

In the task of text-to-molecule retrieval, our model achieves a Hits@1 of 56.5\% and Hits@10 of 94.1\%.
Compared with the previous SOTA model Atomas-base, our model has a commendable improvement of 6.4\% in terms of Hits@1 metric, highlights the effectiveness of ours in aligning molecules and texts.
In the molecule-to-text task, our model also shows superior performance, achieving an improvement of 5.7\% and 1.7\% on Hits@1 and Hits@10 compared with AMAN.

In these two tasks, the value of Mean Rank metric decreases by 1.83 and 2.83, respectively.
Concurrently, the value of MRR metric has an improvement of 0.049 and 0.048.
These results indicate that our model can not only retrieve the matching answer for more instances, but also improve the sorting result of the whole dataset.

\begin{figure}[t]
\centering
\includegraphics[width=1\linewidth]{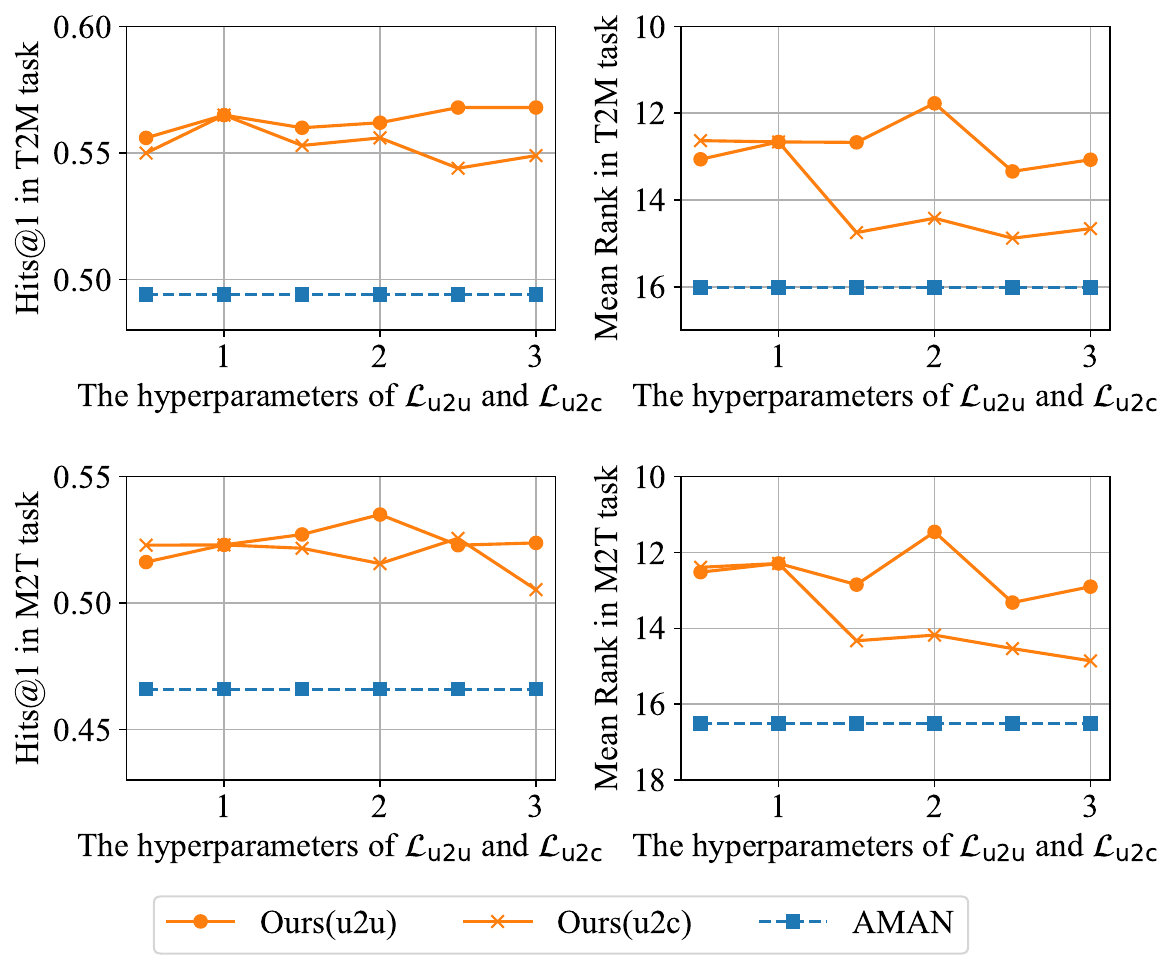}
\DeclareGraphicsExtensions.
% \vspace{-10pt}
\caption{Results of our model w.r.t different weights of $\mathcal{L}_{\mathrm{u2u}}$ and $\mathcal{L}_{\mathrm{u2c}}$ in text-to-molecule (T2M) retrieval task and molecule-to-text (M2T) retrieval task. The variant of our model, denoted as Ours(u2c), pertains to the model when adjusting the weight of $\mathcal{L}{\mathrm{u2c}}$, while Ours(u2u) refers to the model when adjusting the weight of $\mathcal{L}{\mathrm{u2u}}$.}
\label{fig:hyper}
\end{figure}

\subsection{Ablation Study}

To better investigate the effectiveness of
different components in our model, we conduct an ablation study as reported in Table \ref{tab:ablation}.

First, eliminating any loss results in the performance degradation of our model.
From line 1 and line 3, we can observe that removing the entire $\mathcal{L}_{\mathrm{u2u}}$ causes a 2.7\% drop in the Hits@1 metric in the text-to-molecule retrieval task.
Meanwhile, the Mean Rank metric increases by 2.7.
Furthermore, the removal of $\mathcal{L}_{\mathrm{u2c}}$ results in a 3.2\% decline in the Hits@1 metric and a 3.93 increase in Mean Rank metric.
These results demonstrate the validity of the two losses, with $\mathcal{L}_{\mathrm{u2c}}$ enhancing performance more effectively.

Second, we remove the memory bank on the base model without any second similarity loss, leaving only a modality-shared linear layer instead.
The result in line 6 shows that this change causes a significant performance decline in all metrics, showing the effectiveness of these memory vectors.
We conjecture that the modality-shared memory bank provide a more stable and generalizable means to learn multi-modal representations, benefiting contrastive learning.

\subsection{Analysis of Hyperparameters}
\label{hyper}

As described in Equation~\ref{loss}, we do not introduce hyperparameters for both $\mathcal{L}_{\mathrm{u2u}}$ and $\mathcal{L}_{\mathrm{u2c}}$. Intuitively, introducing more hyperparameters may enable them to exert greater impacts.
To this end, we assign two hyperparameters to $\mathcal{L}_{\mathrm{u2u}}$ and $\mathcal{L}_{\mathrm{u2c}}$, respectively, and vary each hyperparameter from 0.5 to 3.0 with an increment of 0.5 while fixing all other hyperparameters, so as to investigate the sensitivity of our model.

We report the Hits@1 and Mean Rank of the previous SOTA model, AMAN, and our model on the ChEBI-20 test set in Figure~\ref{fig:hyper}. The results show that our models always achieves better performance than AMAN, indicating the insensitivity of our model to the hyperparamaters.

\begin{figure}[t]
\centering
\includegraphics[width=0.75\linewidth]{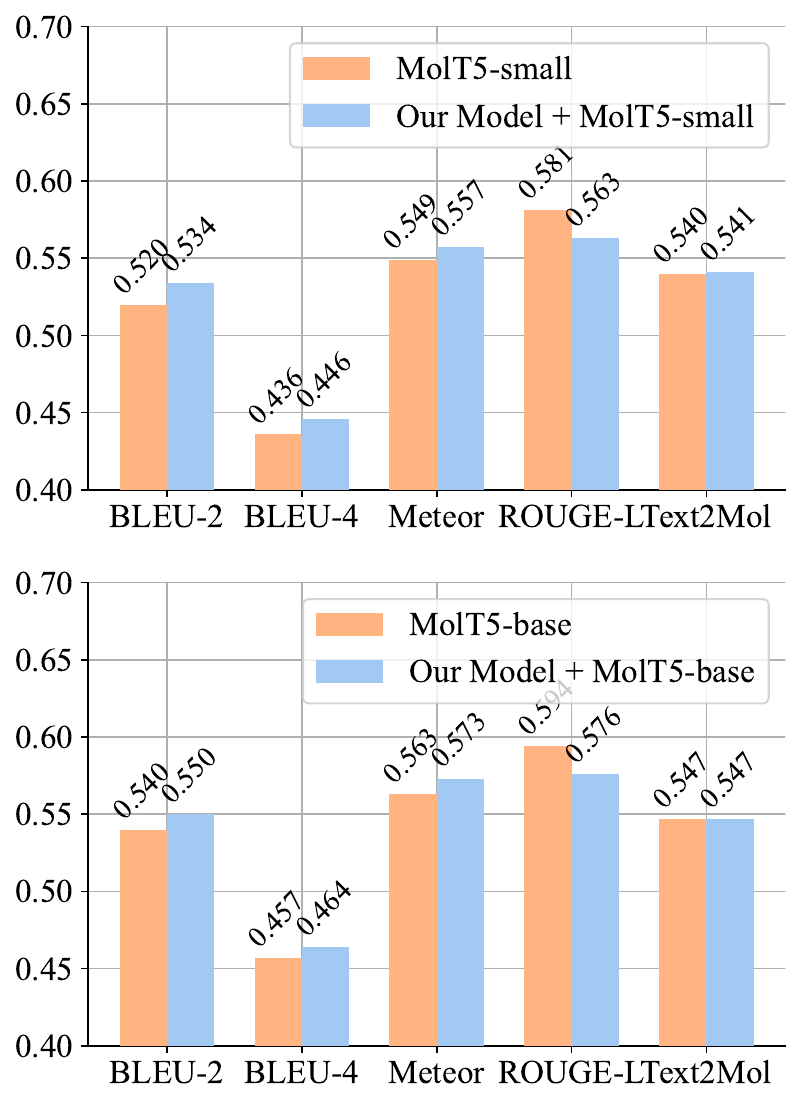}
\DeclareGraphicsExtensions.
% \vspace{-5pt}
\caption{Comparison of MolT5 and our model-enhanced MolT5 on molecule caption task on the ChEBI-20 test set. Our model + MolT5 represents the performance after concatenating molecule graph features to the input of the MolT5 encoder.}
\label{fig:molecule_caption}
\end{figure}

\subsection{Comparison with Pretrain-finetune Paradigm based Models}

We finetune our model on the PCdes dataset \cite{kvplm} and then compare it with the following commonly-used models based on pretrain-finetune paradigm:

\begin{itemize}
\setlength{\itemsep}{2pt}
\setlength{\parsep}{2pt}
\setlength{\parskip}{2pt}
\item \textbf{SciBERT} \cite{scibertBeltagyLC19}. This model is pre-trained on an extensive collection of scientific literature. During the finetuning process, molecules are represented using SMILES sequences and share the same BERT encoder with texts.
\item \textbf{KV-PLM*} \cite{kvplm}. 
This model is firstly initialized with SciBERT and then performs mask language modeling on a new corpus where SMILES sequences are appended after molecule names. 
It employs two distinct tokenizers, with the BPE algorithm utilized for segmenting SMILES sequences and the SciBERT tokenizer employed for segmenting text descriptions.
\item \textbf{MoMu-K} \cite{momusu}. Using GraphCL~\cite{you2020graphcl} as the molecule encoder and KV-PLM~\cite{kvplm} as the text encoder, it is a molecular multi-modal foundation model pre-trained using molecule graphs and their semantically-related text descriptions via contrastive learning.
\end{itemize}

When finetuning baselines and our model, we set the batch size, the number of epochs, and the learning rate to 64, 30, and 5e-5, respectively.
% We run the experiments three times and report the average results.
As implemented in \cite{kvplm,momusu}, we report Hits@1 and Recall@20 at both paragraph- and sentence-level settings. Particularly, we find that there are 3.7\% and 4.3\% test instances appear in our training set and MoMu’s training set, respectively. Therefore, we delete these instances from the test sets and retest the previous SOTA model, MoMu-K, and our model. Other results are directly cited from their corresponding papers.

Table~\ref{tab:pcdes} presents the performance of baselines and our model on the PCdes test set.
Overall, our model surpasses the baselines across most metrics, with the exception of sentence-level Rec@20 when performing the text-to-molecule retrieval.
It is important to note that the baselines have been pre-trained on extensive data. Despite our model being trained solely on the ChEBI-20 dataset, it achieves commendable performance by finetuning on a limited training dataset of PCdes utilizing our training objectives.

\begin{figure}[t]
    \centering
    \includegraphics[width=0.45\textwidth]{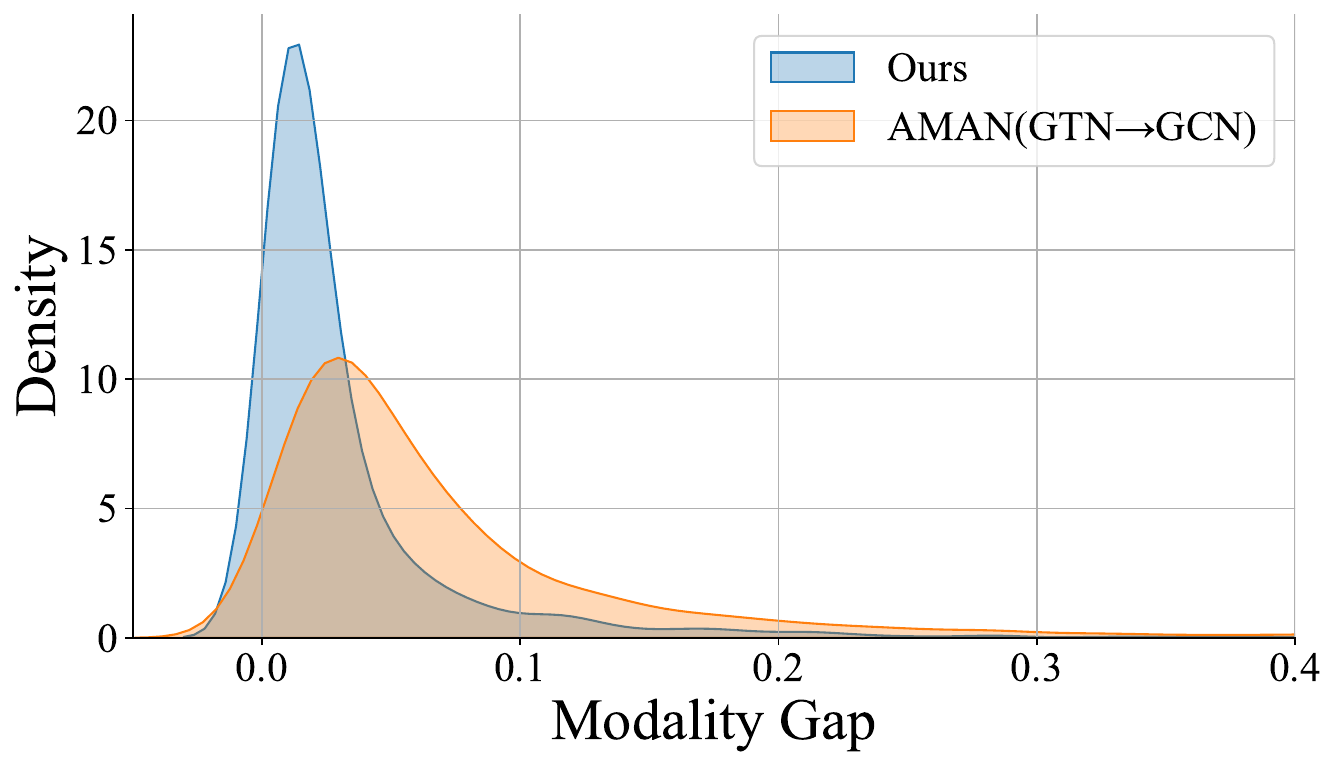}
    % \vspace{-15pt}
    \caption{The distributions of the modality gap with kernel density estimation (KDE) on ChEBI-20 test set.}
    \label{fig:modality_gap}
\end{figure}

\begin{figure*}[t]
\centering
\includegraphics[width=0.85\linewidth]{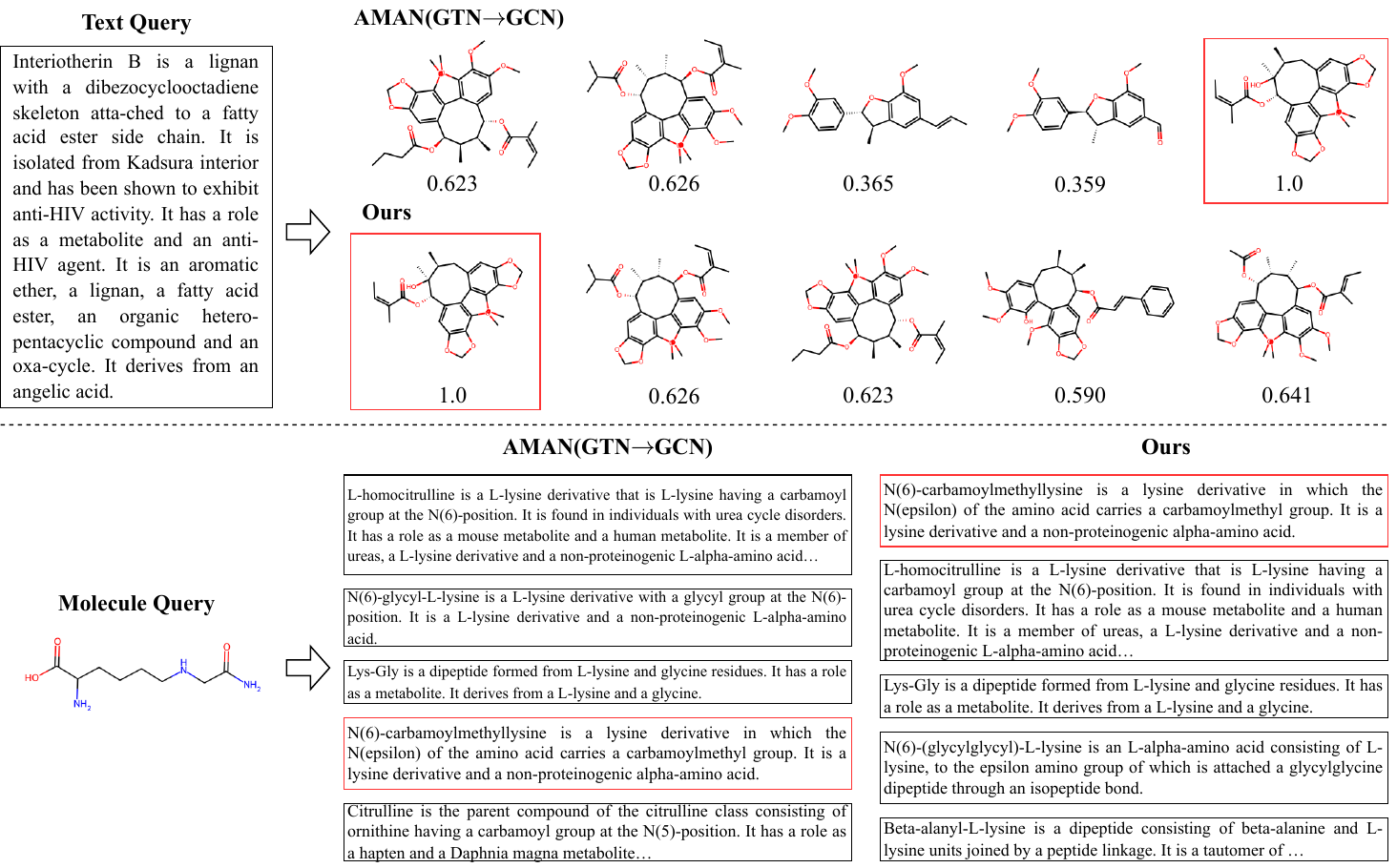}
\DeclareGraphicsExtensions.
% \vspace{-5pt}
\caption{Visualization of top-5 retrieval results of AMAN(GTN$\rightarrow$GCN) and our model. The correctly matched results are boxed in red. In the text-to-molecule retrieval task, each molecule is marked below with its similarity to the ground truth.}
\label{fig:case}
\end{figure*}

\subsection{Molecule Captioning}

Molecule captioning~\cite{molt5} aims to generate a text description for the given molecule.
Following MoMu~\cite{momusu}, we utilize MolT5-small and MolT5-base~\cite{molt5} as baselines to explore whether the molecule representations produced by our model can help generate more accurate text descriptions.
We append the molecule graph representation to the input of the MolT5 encoder through an adapter, which is an MLP.
We only finetune the adapter on the ChEBI-20 dataset, and evaluate MolT5 and our model-enhanced MolT5 with BLEU, METEOR, and Text2Mol metrics.
Results are shown in Figure~\ref{fig:molecule_caption}, our model-enhanced MolT5 achieve better performance in the most metrics, which evidences that the molecule graph representations produced by our multi-modal encoder contain rich and related text information.

\subsection{Analysis}

To validate whether our model successfully learns the modality-shared features, we follow \cite{fang2023cress} to define the \emph{modality gap} of the $i$-th instance as $MG\left(x_i\right)=1-cos\left(\textbf{x}_i^t, \textbf{x}_i^m\right)$.
From Figure~\ref{fig:modality_gap}, we observe that our model has a general decrease in the modality gap compared with AMAN(GTN$\rightarrow$GCN), suggesting that the representations of two modalities learned by our model are more consistent.

Besides, to verify that the similarity between instances does not change with modality, we randomly sample 50,000 instance pairs from the ChEBI-20 test set, and then calculate the average difference between the cosine similarity of the text representations of two instances and that of their molecule representations on the ChEBI-20 test set.
Finally, we obtain 0.091 and 0.051 for AMAN(GTN$\rightarrow$GCN) and our model, respectively.
It indicates that our model learns enhanced modality-shared features from the perspective of similarity between instances.

\subsection{Case Study}

To further validate the effectiveness of our model, we visualize the retrieval results of our model and AMAN(GTN$\rightarrow$GCN), which is our most important baseline.
As illustrated in Figure~\ref{fig:case},
we choose \emph{Interiotherin B} and \emph{N(6)-carbamoylmethyllysine} as the text query and molecule query, respectively.
AMAN(GTN$\rightarrow$GCN) fails to rank the ground truth in the first position, while our model successfully accomplishes the task.

Besides, in the text-to-molecule retrieval task, we compute the similarity between the retrieved molecules and the ground truth based on Morgan Fingerprint \cite{morgan1965generation}.
Our retrieved molecules exhibit higher similarity scores than those of AMAN(GTN$\rightarrow$GCN), indicating that our model effectively captures the semantic relatedness between text and molecules.
\section{Related Work}

Text2Mol \cite{edwards2021text2mol}, which is specifically designed for the text-to-molecule retrieval task, pioneers the integration of text and molecule modalities.
It is equipped with two encoders to encode the two modalities separately, and leverages contrastive learning to align modalities.
Similarly, both MoMu \cite{momusu} and MoleculeSTM \cite{liu2022multimoleculestm} represent molecules as graphs and 
employ the same architecture and training objective.
Subsequent advancements to this approach, referred to as AMAN \cite{zhao2023adversarial} and Atomas~\cite{atomas2024zhang}, achieve the significant performance by introducing adversarial training and multi-grained feature alignments to conduct better modality alignment, respectively.
Besides, some studies~\cite{kvplm, liu2023molxpt} adopt 1D SMILES sequences to represent molecules. They conduct pre-training on the data, where SMILES sequences are inserted into the paired texts, thus facilitating implicit cross-modality alignment.

Despite these advancements, most methods overlook the modality interaction from aspect of the model architecture and only focus on the first-order similarity losses such as contrastive learning loss.
Recently, a common practice in language-image learning is to incorporate a cross-modal interaction module subsequent to encoders.
For instance, ViLBERT \cite{vilbert}, LXMBERT \cite{lxmert}, and Erney-Vil \cite{yu2021ernie} employ a bidirectional cross-attention component for modal fusion, enabling the interaction between the two modalities.
BLIP2 \cite{li2023blip2} and Chimera \cite{chimera} introduce a series of learnable queries to capture modality-shared information.
Besides, as a promising multi-modal approach, similarity-based knowledge distillation \cite{DBLP:journals/npl/SuSLXWC18, DBLP:conf/cvpr/ParkKLC19,mao2022cmd,tran2022within,DBLP:conf/cvpr/WuWZLT22} has been explored to achieve the semantic alignment between two representation spaces from the perspective of the relational feature.

Drawing the inspiration from prior studies, we design not only a novel modality-shared feature projector but also second-order similarity losses to learn better multi-modal representations tailored specifically for molecule-text retrieval tasks.
\section{Conclusion}
In this paper, we propose a cross-modal text-molecule retrieval model that greatly improve previous works in two aspects.
We not only introduce learnable memory queries but also propose second-order similarity losses to further enhance cross-modality alignment.
Experimental results on the ChEBI-20 dataset and PCdes dataset demonstrate that our model outperforms all existing competitive baselines.
In the future, we plan to investigate the utilization of knowledge in large language models to further enhance the performance in cross-modal text-molecule retrieval.

\section*{Acknowledgements}
This work is supported by National Natural Science Foundation of China (Grant No.62262072, 62472370) and Yunnan Provincial Philosophy and Social Science Planning Social Think Tank Project (Grant No.SHZK2024204).

% National Natural Science Foundation of China (No. 62262072).

\bibliographystyle{ieeetr}
\bibliography{ref}

\end{document}